\documentstyle[preprint,eqsecnum,aps,prb]{revtex}

\begin{document}
\title{Investigation of A$_{1g}$ phonons in YBa$_2$Cu$_3$O$_7$ by means 
of LAPW atomic--force calculations}
\author{Robert Kouba and Claudia Ambrosch-Draxl}
\address{Institut f\"{u}r Theoretische Physik, Universit\"{a}t Graz, 
Universit\"{a}tsplatz 5, A--8010 Austria}
\date{\today}
\maketitle

\begin{abstract} 
We report first--principles frozen--phonon calculations for the determination
of the force--free geometry and the dynamical matrix of the five Raman--active
A$_{1g}$ modes in YBa$_2$Cu$_3$O$_{7}$. To establish the shape of the phonon
potentials atomic forces are calculated within the LAPW method. Two different
schemes -- the local density approximation (LDA) and a generalized gradient
approximation (GGA) -- are employed for the treatment of electronic exchange
and correlation effects. We find that in the case of LDA the resulting phonon
frequencies show a deviation from experimental values of approximately -10\%.
Invoking GGA the frequency values are significantly improved and also the
eigenvectors are in very good agreement with experimental findings.
\end{abstract}


\narrowtext

\section{INTRODUCTION}
The phonons in high temperature superconductors have been intensively studied
in the last decade, where Raman--scattering is the most often used experimental
technique. The phonon frequencies of different symmetries are measured for a
variety of high--$T_c$ (HTC) compounds. On the other hand, much less is known
about the corresponding eigenvectors. Experimentally, information is accessible
when certain atoms in the unit cell are replaced by chemically equivalent
elements. By analyzing the frequency shifts from Raman measurements performed
on isotope--substituted samples the composition of the normal modes can be
obtained.  The experimentally best investigated HTC crystal is
YBa$_2$Cu$_3$O$_{7-x}$.  Raman measurements on isotope--substituted samples of
this material have given insight into the coupling of atomic vibrations
especially for the two A$_{1g}$ phonons containing Barium and copper
oscillations.\cite{cardona,henn}

In the past years, ab--initio calculations of frequencies and eigenvectors of
high--symmetry phonons in YBa$_2$Cu$_3$O$_{7}$ have been
published.\cite{cohen,rodriguez} These works were carried out by the
frozen--phonon technique within all--electron band--structure calculations
utilizing the local density approximation (LDA) for the treatment of electron
exchange and correlations. The phonon potentials were derived from
total--energy values for different ionic configurations. The so obtained
A$_{1g}$ phonon frequencies show a tendency of being slightly too soft, lying
approximately 10\% below their experimental counterparts.

This deviance can be discussed in several terms. First, the sole employment of
total--energy values for establishing the dynamical matrix requires
self--consistent electronic--structure calculations for a huge number of atomic
configurations. When only harmonic terms are taken into account this number is
proportional to $N^2$ when $N$ atoms are involved in the vibrations of the same
symmetry class. Thus the number of actually used total energy values hardly
exceeds the minimum number of configurations. Anharmonic contributions to the
phonon potentials which might change the diagonal and off--diagonal elements of
the dynamical matrix could not be accounted for. Second, the strong electron
correlations in high--T$_{c}$ materials are known to be underestimated by LDA
leading to some shortcomings in the description of the electronic density
distribution \cite{efg}. This fact might also bring up certain inaccuracies in
the description of structural and vibronic properties of this material.

In this paper we present frozen--phonon calculations employing the linearized
augmented plane--wave (LAPW) method.\cite{andersen,wien95} We evaluate the
atomic forces acting on the ions when distorted from
equilibrium.\cite{yu,kohler} Using these data instead of total energy values
the geometry is optimized and the phonon potentials of the five A$_{1g}$ modes
are constructed. We checked the necessity of including anharmonic terms in the
description of the energy surface and the number of configurations required for
a stable fit. This procedure was carried out within the local--density
approximation first. Alterations of the phonon potentials when treating
electron exchange and correlations via a generalized gradient approximation
(GGA)\cite{perdew} rather than via LDA are also investigated.

\section{Crystal Structure and Symmetry}
YBa$_2$Cu$_3$O$_7$ is an orthorhombic crystal with the space group Pmmm.
Yttrium and the chain atoms Cu(1) and O(1) occupy sites with the full symmetry
of the cell (mmm).  For barium, the copper and oxygen atoms in the planes --
Cu(2), O(2), and O(3) -- and the bridging oxygen O(4) the site symmetry is
lower (mm), with a vibrational degree of freedom along the c--axis.  Thus the
five $A_{1g}$ Raman--active modes are coupled c--axis vibrations of these
atoms.

\section{METHOD} 
\subsection{Energy Surface} 
Within the frozen--phonon approach potentials underlying lattice vibrations are
traditionally constructed through polynomial fitting of total--energy values
calculated at selected frozen--in distortions of the crystal. If $N$ denotes
the number of coupled degrees of freedom in a certain phonon symmetry, the
minimal number of total--energy calculations necessary for the harmonic
approximation of the potentials scales with $N^2$. This scaling behavior in
conjunction with the formidable computational effort of self--consistent
band--structure calculations for polyatomic structures limits the feasibility
of this approach when applied to YBa$_2$Cu$_3$O$_7$. Thus, in past
publications,\cite{cohen,rodriguez} the number of actually calculated energy
values hardly exceeded the minimal number necessary for the quadratic energy
surface. In this way, neither anharmonic effects in the phonon potentials can
fully be taken into account nor is it possible to perform a reliable check on
the stability of the fitting parameters. These limitations can be overcome by
evaluating the forces acting on atoms when they are displaced from
equilibrium. Atomic--force formalisms\cite{yu,pseudo} have proved to give an
accurate description of the total energy gradient with respect to ionic
positions and thus enlarge the information on the shape of the energy surface
obtainable from each frozen--phonon calculation. In fact, for the harmonic
approximation it suffices to consider a number of distortion patterns scaling
with $N$ rather than $N^2$.  The additional numerical effort required for the
evaluation of atomic forces is not significant.

\subsection{Atomic Forces} 
If the electronic density distribution $\rho({\bf r})$ in a crystal is
precisely known, the Hellmann--Feynman theorem\cite{hellmann,feynman} states
that the atomic force ${\bf F}^{at}_{\alpha}$ (defined as the negative
derivative of the total energy $E_{tot}$ with respect to the nuclear coordinate
${\bf R}_{\alpha}$) is exactly described by the electrostatic force exerted on
the nucleus $\alpha$ by all other charges of the system:

\begin{eqnarray} 
{\bf F}_{\alpha}^{at} & \equiv &  -\frac{dE_{tot}}{d\bf R_{\alpha}} =
- q_{\alpha} \sum_{\beta \not= \alpha} q_\beta\frac{({\bf R}_{\alpha} - 
{\bf R}_{\beta})}{\mid {\bf R}_{\alpha} - {\bf R}_{\beta} \mid^3 }
+q_\alpha \int_{\Omega} \rho({\bf r}) \frac{({\bf R}_{\alpha} - {\bf r})}
{\mid {\bf R}_{\alpha} - {\bf r} \mid^3} d^3 {\bf r}
\end{eqnarray} 

\noindent with $q_{\beta}$ denoting the nuclear charge at site {\bf R}$_\beta$.
This electrostatic force is called Hellman--Feynman force and is denoted ${\bf
F}_{\alpha}^{HF}$ in the following.  In practical ab--initio calculations
charge densities are never exact and so correction forces have to be added to
the HF--force to obtain an accurate description of the total--energy
gradient\cite{pulay}. The calculation of these forces requires an analytic
treatment of the first--order change of the energy expression when the ionic
positions ${\bf R}_{\alpha}$ are displaced by a small amount $\delta{\bf
R}_{\alpha}$. The energy expressions depend on the approximations made when
simplifying the many--body crystal Hamiltonian.  In the case of the
all--electron band--structure method LAPW,\cite{andersen} which is based on
density--functional theory (DFT) and a local exchange--correlation potential,
this results in two corrective force--terms so that the atomic force can be
written as

\begin{eqnarray} 
\label{eq_Ftot}
{\bf F}_{\alpha}^{at}=
{\bf F}_{\alpha}^{HF}+{\bf F}_{\alpha}^{core}+{\bf F}_{\alpha}^{IBS}
\end{eqnarray} 

\noindent 
${\bf F}^{core}_{\alpha}$ is due to the simplifications made for the
core elctrons' wave--functions. The core--states are assumed to be
non--dispersive in ${\bf k}$--space and are obtained by solving the
Schr{\"o}dinger equation neglecting the non--spherical part of the effective
crystal potential $V_{eff}({\bf r})$. It was shown in
Ref.\onlinecite{yu} that ${\bf F}^{core}_{\alpha}$ can explicitly be calculated by

\begin{eqnarray} 
{\bf F}_{\alpha}^{core}= - \int_{\Omega} \rho_{core}^\alpha(r) \nabla
V_{eff}({\bf r})d^3 {\bf r}
\end{eqnarray}

\noindent ${\bf F}^{IBS}_{\alpha}$ is related to the fact that only a finite
number of basis functions can be taken into account when solving the eigenvalue
problem for the valence electrons by a variational scheme. In our particular
case, the wave--functions are linear combinations of {\it linearized augmented
plane waves} (LAPWs). Since these LAPWs depend on the atomic positions the
incompleteness of the set of basis functions (IBS) leads to one part of the
correction force ${\bf F}^{IBS}_{\alpha}$. A further contribution termed ${\bf
D}_{\alpha}$ stems from the discontinuity of the second derivative of the
wave--functions at the atomic sphere boundaries. The total ${\bf F}^{IBS}$ is
given by\cite{yu}

\begin{eqnarray}
{\bf F}_{\alpha}^{IBS} & = &
- \sum_{i}  w_i \left[ \langle \frac{d \psi_{i}}
{d {\bf R}_{\alpha}} | \hat T+V_{eff} 
- \varepsilon_{i} | \psi_{i} \rangle
+ \langle \psi_{i} | \hat T+V_{eff} - \varepsilon_{i} | 
\frac {d \psi_{i}}{d{\bf R}_{\alpha}} \rangle \right] 
- {\bf D}_{\alpha} \nonumber \\
\end{eqnarray}
\noindent with
\begin{eqnarray}
{\bf D}_{\alpha} & = & \sum_i w_i {\oint_{{MT}_{\alpha}}  
\Bigl [\psi_{i}^*({\bf r}) \hat T
\psi_{i}({\bf r})|_{MT_{\alpha}} - 
\psi_{i}^*({\bf r}) \hat T 
\psi_{i}({\bf r})|_{Int} \Bigr ] d {\bf A}_{\alpha} }.
\end{eqnarray}

\noindent $\psi_{i}$ denotes the wavefunction of a valence--state, $w_i$ the
respective occupation number, and $\varepsilon_{i}$ the corresponding
eigenvalue. The integration in the term $\bf D_{\alpha}$ runs over the surface
of the atomic sphere around the nucleus $\alpha$.  $\psi_{i}({\bf
r})|_{MT_{\alpha}}$ is the eigenstate expressed in terms of the local basis
functions within the atomic sphere. $\psi_{i}({\bf r})|_{int}$ is the
interstitial part of the wave--function where the basis set consists of
plane--waves.

\section{Calculations} 
\subsection{Energy Surface and Dynamical Matrix}
The zone--center $A_{1g}$ Raman--active modes in YBa$_2$Cu$_3$O$_7$ are coupled
c--axis vibrations of Ba, Cu(2), O(2), O(3), and O(4). Using experimentally
determined lattice constants\cite{beno} (a~=~3.8231~\AA, b~=~3.8864~\AA,
c~=~11.6807~\AA) atomic forces acting on these atoms were calculated for 19
different distortion patterns in the vicinity of the force--free geometry. 10
patterns corresponded to single--atom displacements, in the remaining nine
configurations two atoms were distorted simultaneously. The maximum amplitude
of displacement from equilibrium was approximately 0.05 \AA. This yielded a set
of 95 values for establishing a least square fit of the energy surface with
respect to the five coupled degrees of freedom. To optimize this fit anharmonic
coefficients supplemented the linear and quadratic terms where necessary.  On
the basis of the 15 independent harmonic force--constants of the resulting
polynomial function we set up the dynamical matrix for calculating phonon
frequencies and eigenvectors. Additionally, the stability of the fit parameters
was tested by repeating the fitting procedure on the sole basis of the 10
single--atom displacement patterns. The resulting change in the phonon
frequencies was very small (less than $\pm 2$ $cm^{-1}$) indicating a high 
reliability of the polynomial coefficients.

\subsection{Computational Details}
The self--consistent band--structure calculations for each frozen--in
distortion were done using the full--potential LAPW method as implemented in
the WIEN95--code.\cite{wien95} Valence states and semicore states (Y--4s,
Y--4p, Ba--5s, Cu--3p) were treated in the same energy window by using local
orbitals\cite{lo} as an extension to the LAPW basis set.  The expansion of the
wave functions included 1950 LAPW's ($RK_{max}=7.1$). The sphere radii were
chosen to be 2.5 a.u. for Y and Ba, 1.8 a.u. for Cu(1), 1.9 a.u. for Cu(2) and
1.55 a.u. for O(1), O(2), O(3) and O(4). Charge densities and potentials in the
atomic spheres were represented by spherical harmonics up to $L=6$, whereas in
the interstitial region these quantities were expanded in a Fourier series with
~3500 stars of ${\bf K}$ in the LDA case and ~5000 stars of ${\bf K}$ in the
GGA case. The radial functions of each LAPW were calculated up to $l=12$ and
the non--spherical potential contribution to the Hamilton matrix (Gaunt
coefficients) had an upper limit of $l=6$. For Brillouin--zone (BZ)
integrations a $12 \times 12 \times 4$ ${\bf k}$--point mesh was used yielding
72 ${\bf k}$--points in the irreducible wedge.

\subsection{Correction Terms}
The unreliability of the pure HF--force for calculating the atomic force within
the LAPW method has been demonstrated by Yu et al.\cite{yu} for phonons in
simple systems as Mo and Si. In the case of YBa$_2$Cu$_3$O$_7$ we can confirm
the importance of the corrective force terms for an accurate description of the
total--energy gradient. In Table \ref{t_contributions} we list the various
contributions to the atomic forces when the two atoms O(2) and O(4) are
displaced simultaneously from equilibrium. With the exception of O(2), ${\bf
F}^{IBS}$ and ${\bf F}^{core}$ are of the same magnitude as ${\bf F}^{HF}$ but
of different signs. Due to this cancellation the resulting total force
(\ref{eq_Ftot}) is one or even two orders of magnitude smaller than its
constituents. Thus a very precise calculation is required in order to obtain
reliable results.

\section{Results}
\subsection{LDA Results}
For the case, where the electronic exchange and correlation effects are treated
by LDA, the energy surface is well described by harmonic terms plus a cubic
term in the O(4) coordinate.  The minimum energy positions of this polynomial
function are $z_{Ba}~=~0.1812$~c, $z_{Cu(2)}~=~0.3530$~c,
$z_{O(2)}~=~0.3789$~c, $z_{O(3)}~=~0.3783$~c, and $z_{O(4)}~=~0.1586$~c, which
agree well with the experimental structure parameters ($z_{Ba} = 0.1843$~c,
$z_{Cu(2)}~=~0.3556$~c, $z_{O(2)}~=~0.3772$~c, $z_{O(3)}~=~0.3789$~c, and
$z_{O(4)}~=~0.1584$~c). The eigenvalues and eigenvectors of the corresponding
dynamical matrix are displayed in Table \ref{t_LDA}. The two low--lying modes
are governed by hybridized Ba--Cu motions. Further we observe an appreciable
admixture of O(2), O(3) and O(4) oscillations in the 330, 387 and 452 $cm^{-1}$
modes. The importance of the anharmonic term in the O(4) coordinate is
reflected in the fact, that its exclusion from the fitting procedure leads to a
lowering of the corresponding apex mode (452 $cm^{-1}$) by 8 $cm^{-1}$. The
introduction of other diagonal or off--diagonal anharmonic terms had the
maximum effect of $\pm1$ $cm^{-1}$ on the phonon frequencies.

In the comparison of the calculated phonon--frequencies with their experimental
counterparts it can be noticed that the discrepancies lie in the range of
$-10\%$. The only exception is the out--of--phase motion of O(2) and O(3) where
the agreement with the experimental frequency is excellent. These findings are
close to the results of Rodriguez et al.\cite{rodriguez} based on LMTO
total--energy calculations but reveal larger differences to LAPW frozen--phonon
calculations by Cohen et al.\cite{cohen} particularly for the three modes
dominated by the motions of the oxygens. They find a frequency of 513 $cm^{-1}$
for the apex mode, 12\% above our calculated value. For the O(2)-O(3) mode and
the O(2)+O(3) mode their values are 312~$cm^{-1}$ and 361~$cm^{-1}$, whereas
our results are 335~$cm^{-1}$ and 387~$cm^{-1}$, respectively.  As little
information is given on the computational details in the paper of Cohen et
al.\cite{cohen} we can not give a definit explanation for the discrepancies. We
can state, that the influence of anharmonicities neither on the diagonal nor on
the off--diagonal elements of the dynamical matrix is strong enough to account
for the different values. Parts of the differences might be connected to the
fact, that Cohen et. al used experimental equilibrium positions to optimize
each coordinate separately. It should be noted that in a recent refinement
\cite{pickett} of their work the frequency of the apex oxygen mode is in much
better agreement with our results.

\subsection{GGA Results and Discussion} 
It is well known that the correlation effects between the electrons in
Y-Ba-Cu-O compounds seem to play an important role for the electronic properties
of these materials: Measurements of the electric--field gradient in
YBa$_2$Cu$_3$O$_7$ have revealed that band--structure calculations on the basis
of LDA show certain inaccuracies in the description of the materials'
electronic charge distribution.\cite{efg} In addition, such calculations fail
completely in reproducing the insulating and antiferromagnetic ground state of
YBa$_2$Cu$_3$O$_6$.\cite{ambrosch} This raises the question if correlation
effects also influence the structural properties of YBa$_2$Cu$_3$O$_7$ and thus
might account for the theoretical underestimation of the A$_{1g}$ phonon
frequencies. To obtain an insight into the shortcomings of LDA concerning
vibronic properties, we used a generalized gradient approximation proposed by
Perdew et al.\cite{perdew} for the treatment of the exchange--correlation
energy and potential which is now calculated from the local electronic density
as well as from its gradient. We established the energy surface of the A$_{1g}$
phonons following the same procedure as in the LDA case.

We find that the GGA scheme enhances the diagonal terms in the dynamical matrix
by approximately 15\% (Ba), 10\% (Cu), 3\% (O(2)), 5\% (O(3)), and 4\% (O(4)),
respectively in comparison to the LDA results. On the other hand, the coupling
between the copper and the three oxygen atoms is significantly decreased. The
same applies to the force--constants coupling the apex vibration to the motion
of the other two oxygens. The GGA approach neither changes the anharmonic cubic
term in the O(4) coordinate noticeably, nor does it cause further
anharmonicities in the energy surface. Resulting phonon frequencies and
eigenvectors are displayed in Table~\ref{t_GGA}. We observe a significant
improvement for the Ba and Cu dominated modes, bringing their frequency values
in excellent agreement with their experimental counterparts and also
reproducing the mode admixture very accurately. From isotopic substitution
experiments on the copper--sites the values for all three independent elements
of the dynamical matrix of the Ba--Cu subsystem could be
extracted.\cite{cardona} It can be observed (Table~\ref{t_force}) that the LDA
calculations underestimate both diagonal terms by 15\% whereas the
off--diagonal term lies within the experimental uncertainty.  The respective
values from GGA calculations exhibit perfect agreement with the experiment.

With regard to the oxygen modes, we find that the frequency of the in--phase
O(2)+O(3) mode is increased ranging approximately 8\% below the experimental
value, whereas the phonon frequencies of the out--of--phase O(2)-O(3) mode
remains basically unaltered accurately reflecting the experimental value. The
increase in the diagonal force--constant of the apex oxygen is not accompanied
by a corresponding increase in the frequency of the O(4)--mode. This is due to
the changes in the coupling between O(4) and the other atoms. Concerning the
mode admixtures, no experimental data are available to fully determine the
off--diagonal elements of the O(2)--O(3)--O(4) subsystem. Nevertheless,
site--selective isotope substitution experiments\cite{kaldis} indicate that the
apex mode is largely dominated by the O(4) vibration and that the hybridized
O(2)+O(3) mode has a small but noticeable admixture of the apex oscillation. On
contrary, the O(2)-O(3) mode is decoupled from the O(4) motion. Our
calculations confirm this picture.

\section{Conclusions}

We have determined phonon frequencies and eigenvectors of the five
Raman--active A$_{1g}$ modes in YBa$_2$Cu$_3$O$_{7}$ employing frozen--phonon
calculations on the basis of the full--potential LAPW method. Atomic--force
values were used to establish the atomic equilibrium positions and to map out
the phonon potentials. When treating the electronic exchange and correlations
by LDA, the calculated phonon frequencies display deviations from the measured
values by approximately 10\%. The only exception is the out--of--phase
O(2)-O(3) mode where the agreement with experiment is excellent. Going beyond
LDA by invoking a GGA exchange--correlation functional leads to an overall
improvement of the frequencies. The two low--lying Ba--Cu modes as well as the
O(2)-O(3) mode are in perfect agreement with experimental values whereas the
frequency of the O(2)+O(3) mode is enhanced but still 8\% too small. In
comparison to LDA calculations only the O(4) frequency remains unaltered
ranging 10\% below its measured counterpart. Regarding the mode admixtures we
find that our GGA results for the two Ba--Cu modes very accurately reproduce
the diagonal and off--diagonal force--constants which have been determined by
isotopic substitution experiments. Additionally, our calculations show that the
O(2)-O(3) mode is completely decoupled from the O(4) oscillation, whereas there
is a finite contribution of the O(4) vibration to the O(2)+O(3) mode. The
domination of the O(4) displacement in the apex mode is also revealed. These
findings are compatible with Raman measurements performed on site--selective
substituted samples. In summary, these results indicate that first--principles
calculations based on density functional theory are an appropriate tool for the
investigation of structural and vibronic properties of HTC cuprates, where the
treatment of exchange and correlation effects is an important issue for this
task.

\begin{table}[ht]
\caption[]{Force contributions in mRy for a mixed distortion (-O(2), -O(4)).}
\label{t_contributions}  
\begin{tabular}{lccccc}
  & Ba & Cu(2) & O(2) & O(3) & O(4) \\ 
\hline
position [c]     & 0.1815 & 0.3530 &  0.3740 &  0.3787 &  0.1540 \\ \hline
${\bf F}^{HF}  $ &  46.66 & 123.82 &  0.76   & -8.77   &  290.52 \\
${\bf F}^{core}$ & -36.08 & -88.30 &  6.45   &  0.75   & -188.87 \\
${\bf F}^{IBS} $ & -13.03 & -35.22 &  6.50   &  7.91   &  -75.75 \\ \hline
${\bf F}^{at}  $ & -2.45  &   0.30 & 13.71   & -0.11   &   25.90 \\
\end{tabular} 
\end{table}

\begin{table}[ht]
\caption[]{LDA phonon frequencies in $cm^{-1}$ and eigenvectors of the five
A$_{1g}$ modes in YBa$_2$Cu$_3$O$_7$ at $q=0$ 
compared to experimental data\cite{cardona,burns}.}
\label{t_LDA}  
\begin{tabular}{cccccccc}
{Exp.} & {Theory} & {Rel. Deviation} & {Ba} & {Cu(2)} & {O(2)} & {O(3)} & {O(4)} \\ 
\hline 
116,118 & 103 & -11\%,-13\% & 0.85 & 0.52 & 0.05 & 0.05 & 0.00  \\
150,145 & 130 & -13\%,-10\% & 0.53 &-0.84 &-0.08 &-0.07 & 0.06  \\
335     & 327 &  - 2\%      & 0.00 & 0.02 &-0.81 & 0.59 &-0.05  \\
440     & 387 &  -12\%      & 0.02 & 0.09 &-0.51 &-0.74 &-0.43  \\
500     & 452 &  - 9\%      & 0.03 &-0.11 & 0.28 & 0.32 &-0.90  \\  
\end{tabular} 
\end{table}

\begin{table}[ht]
\caption[]{GGA phonon frequencies in $cm^{-1}$ and eigenvectors of the five
A$_{1g}$ modes in YBa$_2$Cu$_3$O$_7$ at $q=0$ 
compared to experimental data\cite{cardona,burns}.}
\label{t_GGA}  
\begin{tabular}{cccccccc}
{Exp.} & {Theory} & {Rel. Deviation} & {Ba} & {Cu(2)} & {O(2)} & {O(3)} & {O(4)} \\ 
\hline 
116,118 & 115 & -1\%,-3\% & 0.91 &  0.40 &  0.05 &  0.02 &  0.02  \\  
150,145 & 144 & -4\%,-1\% & 0.41 & -0.91 & -0.06 & -0.04 &  0.03  \\
335     & 328 &  - 2\%    & 0.01 &  0.04 & -0.81 &  0.60 & -0.02  \\
440     & 405 &  - 8\%    & 0.02 &  0.07 & -0.56 & -0.77 & -0.30  \\
500     & 452 &  - 9\%    & 0.02 & -0.04 &  0.20 &  0.23 & -0.95  \\  
\end{tabular} 
\end{table}

\begin{table}[]
\caption[]{Force constants in N/m for the sublattice vibrations of Cu and Ba
compared to measured values\cite{cardona}. 
k$_{11}$ and k$_{22}$ denote the diagonal force constants, k$_{12}$ the coupling 
term.}
\label{t_force}  
\begin{tabular}{cccc}
  & LDA  & GGA & Exp. \\ 
\hline
k$_{11}$  &  101  & 119  &  118.6 (-5.3/+5.0)    \\
k$_{22}$  &  69   &  77  &   81.3 (-2.2/+2.5)    \\ 
k$_{12}$  & -17   & -15  &  -16.3 (-3.3/+5.6)    \\
\end{tabular} 
\end{table}

\bigskip
\noindent
{\bf Acknowledgments} \newline
This work was supported by the Austrian Science Foundation, project nrs.
P9908-PHY and P11893-PHY.


\begin{references}

\bibitem{cardona} T. Strach, T. Ruf, E. Sch\"onherr, and M. Cardona, Phys. Rev.
B {\bf 51}, 16460 (1995).

\bibitem{henn} R. Henn, T. Strach, E. Sch\"onherr, and M. Cardona, Phys. Rev.
B {\bf 55}, 3285 (1997).

\bibitem{cohen} R. E. Cohen, W. E. Pickett, and H. Krakauer, Phys. Rev. Lett.
{\bf 64}, 2575 (1990).  

\bibitem{rodriguez} C. O. Rodriguez, A. I. Liechtenstein, I. I. Mazin, O. Jepsen, 
O. K. Andersen, and M. Methfessel, Phys. Rev. B {\bf 42}, 2692 (1990).

\bibitem{efg} K. Schwarz, C. Ambrosch--Draxl,  and P. Blaha, Phys. Rev. B {\bf
42}, 2051 (1990).

\bibitem{andersen} O. K. Andersen, Phys. Rev. B {\bf 12}, 3060 (1975).

\bibitem{wien95} P. Blaha, K. Schwarz, P. Dufek and R. Augustyn,
WIEN95, improved and updated Unix version of the copyrighted WIEN--code,
published by P. Blaha, K. Schwarz, P. Sorantin, and S. B. Trickey, in Comp.
Phys. Commun. {\bf 59}, 399 (1990).  

\bibitem{yu} R. Yu, D. Singh, and H. Krakauer, Phys. Rev. B {\bf 43}, 6411 (1991).  

\bibitem{kohler} B. Kohler, S. Wilke, M.  Scheffler, R. Kouba, and C.
Ambrosch--Draxl, Comp. Phys. Commun. {\bf 94}, 31 (1996).

\bibitem{perdew} J. P. Perdew, J. A. Chevary, S. H. Vosko, K. A. Jackson, M. R.
Pederson, D. J. Singh, and C. Fiolhais, Phys. Rev. B {\bf 46}, 6671 (1992).

\bibitem{pseudo} J. Ihm, A. Zunger, and M. L. Cohen, J. Phys. C {\bf 12}, 4409 (1979).

\bibitem{hellmann} H. Hellmann, {\it Einf{\"u}hrung in die Qunantenchemie}, Deuicke, 
Leipzig, 1937, p. 285.

\bibitem{feynman} R. P. Feynman, Phys. Rev. B {\bf 56}, 340 (1939).

\bibitem{pulay} P. Pulay, Mol. Phys. {\bf 17}, 197 (1069).

\bibitem{beno} M. A. Beno, L. Soderholm, D. W. Capone, D. G. Hinks, J. D.
Jorgensen, I. K. Schuller, C. U. Segre, K. Zhang, and J. D. Grace, 
Appl. Phys. Lett. {\bf 51}, 57 (1987).  

\bibitem{lo} D. Singh, Phys. Rev. B {\bf 43}, 6388 (1991).

\bibitem{burns} G. Burns, F. H. Dacol, F. Holtzberg, and D. L. Kaiser, 
 Solid State Commun. {\bf 66}, 217 (1988).

\bibitem{pickett} W. E. Pickett, Phys. Rev. B (preprint).

\bibitem{ambrosch} C. Ambrosch-Draxl and K. Schwarz,
 Solid State Commun. {\bf 77}, 45 (1991). 

\bibitem{kaldis} D. Zech, H. Keller, K. Conder, E. Kaldis, E. Liarokapis, 
 N. Poulakis, and K. A. M\"uller, Nature {\bf 371}, 681 (1994).

\end{references}
\end{document}